\documentclass[journal,onecolumn,final,12pt]{IEEEtran} 

\usepackage{amsmath,amssymb,amsfonts}
\usepackage{algorithmic}
\usepackage{graphicx}
\usepackage[T1]{fontenc}
\usepackage{balance}
\usepackage{soul}
\usepackage{multicol}
\usepackage{multirow}
\usepackage{array}
\usepackage{url}
\usepackage{makecell}
\usepackage{setspace}
\graphicspath{{./figures/}}
\usepackage{caption}
\usepackage{subcaption}
\usepackage{mathrsfs}
\usepackage{booktabs}

\usepackage[backend=bibtex,style=numeric,sorting=none]{biblatex}

\usepackage[acronym]{glossaries}
\newacronym{mci}{MCI}{Mild cognitive impairment}
\newacronym{eeg}{EEG}{Electroencephalography}
\newacronym{mmse}{MMSE}{Mini-Mental State Examination}
\newacronym{bci}{BCI}{Brain-Computer Interface}
\newacronym{nfb}{NFB}{neurofeedback}
\newacronym{wban}{WBAN}{wireless body area network}
\newacronym{ioth}{IoT-Health}{Internet-of-Things for healthcare}
\newacronym{dbms}{DBMS}{Data Base Management System}
\newacronym{snr}{SNR}{signal-to-noise ratio}
\newacronym{dsmv}{DSM-V}{Diagnostic and Statistical Manual of Mental Disorders}
\newacronym{ai}{AI}{artificial intelligence}
\newacronym{xai}{XAI}{explainable AI}
\newacronym{ml}{ML}{machine learning}

\bstctlcite{IEEEexample:BSTcontrol}

\begin{document}

\spacing{1.15}

\title{ACTA: A Mobile-Health Solution for Integrated Nudge-Neurofeedback Training for Senior Citizens}

\author{Giulia~Cisotto, Andrea~Trentini, Italo~Zoppis, Alessio~Zanga, Sara~Manzoni, Giada~Pietrabissa, Anna~Guerrini~Usubini, and Gianluca~Castelnuovo%
\thanks{\small{G. Cisotto is with the Department of Information Engineering, University of Padova, Padova, Italy, the National Centre for Neurology and Psychiatry, Tokyo, Japan, and the National Inter-University Consortium for Telecommunications (CNIT), Italy (email: giulia.cisotto.1@unipd.it); A. Trentini is with the Department of Computer Science, University of Milan, Italy (email: andrea.trentini@unimi.it). I. Zoppis, A. Zanga, and S. Manzoni are with the Department of Computer Science, University of Milano-Bicocca, Milano, Italy (emails: italo.zoppis@unimib.it; a.zanga3@campus.unimib.it, sara.manzoni@unimib.it). G. Pietrabissa, A. Guerrini Usubini, and G. Castelnuovo are with the Istituto Auxologico Italiano IRCCS, Psychology Research Laboratory, Ospedale San Giuseppe, Verbania, Italy and the Department of Psychology, Catholic University of Milan, Milan, Italy (emails: \{gianluca.castelnuovo, giada.pietrabissa, anna.guerriniusubini\}@unicatt.it).}}
}

\maketitle


\begin{abstract}
As the worldwide population gets increasingly aged, in-home tele-medicine and mobile-health solutions represent promising services to promote active and independent aging and to contribute to a paradigm shift towards a patient-centric healthcare.
In this work, we present ACTA (Advanced Cognitive Training for Aging), a prototypal mobile-health solution to provide advanced cognitive training for senior citizens with mild cognitive impairments,
We disclose here the conceptualization of ACTA as the integration of two promising rehabilitation strategies: the "Nudge theory", from the cognitive domain, and the neurofeedback, from the neuroscience domain.
Moreover, in ACTA we exploit the most advanced machine learning techniques to deliver customized and fully adaptive support to the elderly, while training in an ecological environment.
ACTA represents the next-step beyond SENIOR, an earlier mobile-health project for cognitive training based on Nudge theory, currently ongoing in Lombardy Region.
Beyond SENIOR, ACTA represents a highly-usable, accessible, low-cost, new-generation mobile-health solution to promote independent aging and effective motor-cognitive training support, while empowering the elderly in their own aging.
\end{abstract}

\spacing{1.15}


\normalsize

\section{Introduction}
%
%
%
%

As the worldwide population gets increasingly aged, tele-medicine and mobile-health solutions are becoming key services to promote active and independent aging, and to contribute to a paradigm shift towards a patient-centric healthcare.
In many countries, especially Italy, Portugal in Europe, and Japan in Asia, average population age is rapidly increasing and projections indicate $79\%$ of it will be over $60$ by $2050$, according to the $2017$ report of the United Nations \cite{UN2017}.
\gls{mci} is rapidly becoming one of the most common clinical manifestations affecting the elderly. It is characterized by deterioration of memory, attention, and cognitive function that is beyond what is expected based on age and educational level. \gls{mci} does not interfere significantly with individuals' daily activities. It can act as a transitional level towards dementia with a range of conversion of 10\%-15\% per year. It is crucial to protect older people against \gls{mci} and developing dementia. 
As originally conceived, Petersen defined \gls{mci} by the following criteria: (1) a subjective complaint of a memory disturbance (preferably supported by an informant); (2) objective evidence of a memory deficit; (3) generally preserved cognitive functions; (4) intact activities of daily living; and (5) the absence of dementia \cite{Petersen1999}.

Preventive interventions, e.g., mental activity, physical exercises and social engagement, may help decrease the risk of further cognitive decline \cite{Mendoza2018, DYang2019,Fragile2020}. 
To this aim, "SENIOR - SystEm of Nudge theory based ICT applications for OldeR citizen" (ID 2018-NAZ-0129, rif. 2018-0826), a three-year project funded by CARIPLO Fundation in 2018, currently ongoing at the University of Milano-Bicocca and I.R.C.C.S. San Giuseppe Auxologico of Milan, is developing a new "Nudge theory"-based ICT coach system for monitoring and empowering elderly with \gls{mci} \cite{SENIOR2019}: it aims to collect physiological, psychological and behavioral data in order to provide them personalized advices, suggestions for social engagement and overall wellness, according to the well-known "Nudge theory". "Nudge Theory” \cite{Amin2016} refers to an extensive sociological framework for helping people to take decisions in a smooth way. Through proactive and positive suggestions (i.e., the nudges), "Nudge theory" based cognitive training can help the patients in remembering or focusing attention onto specific tasks. In SENIOR, nudges are conveyed via texts, speeches, or acoustic reminders. A dedicated app has been developed to correctly performing the training.

More recently, degeneration in \gls{mci} has been also addressed by \gls{nfb} interventions \cite{Li2020, Jiray2019, Jiang2017, Schiatti2018}. \gls{nfb} is a strategy to obtain a self-regulation of specific neural substrates which, in turn, produces improvements of specific cognitive and motor functions \cite{Lubar2003}.
The target neural activity (assumed to be pathological) is associated with an automatic output (e.g., a cursor moving on a computer screen), the patient is instructed to control the output and a positive reinforcement can be provided him/her to help them reach their goal: in fact, after some trials-and-errors, the patient learns to control the output. \gls{nfb} is enabled by the basic principles of "operant learning" and "brain plasticity" \cite{Cisotto2014,Silvoni2011}. Operant learning is employed to create an association between brain activity and (cognitive) behaviour in real-time, and facilitates the learning process via reinforcement strategy. Brain plasticity, instead, is the key ability of the brain to adapt its own functionalities to finally control the output.
While criticisms still persist on the mechanisms through which \gls{nfb} may produce benefits to specific patients, an increasing bulk of evidence has proved it as effective to improve motor and cognitive (e.g., attention deficits, memory, cognitive flexibility, reaction time, and executive functions) functionalities in a variety of pathologies \cite{Lubar2003, Frontiers2014, Sung2012, Belkacem2020}, including \gls{mci} \cite{Lavy2019, Marlats2020, Monaco2019,Silva2019}. 

The abovementioned studies used \gls{eeg}, a non-invasive electrophysiological technique that allows to acquire the individual's brain activity with a very high time resolution (i.e., milliseconds), making it ideal to deliver timely \gls{nfb} reinforcing stimuli \cite{Book2014, Book2015, ICC2013, WVITAE2014, Healthcom2013}.
\gls{eeg}-based \gls{nfb} is typically operated by \gls{bci}, i.e., "closed-loop" human-machine systems that process \gls{eeg} through several steps of pre-processing, feature extraction and classification, and transform the classifier's output in a command for an external device, e.g., a cursor on a computer or a robotic device navigating a room. During training, the machine refines its model of the user's brain activity and, at the same time, the human learns how to modulate his/her own brain activity to control the machine's output \cite{Silvoni2011, Frontiers2014}.
Interestingly, \cite{Lavy2019}, the authors showed that central alpha \gls{eeg} \gls{nfb} was able to improve memory performance in \gls{mci} patients, and this outcome has been maintained at 30-day follow-up. Beta band EEG NFB was also reported to produce significant improvements to MCI patients in domains such as attention, memory, cognitive flexibility, reaction time, and executive functions \cite{Jiang2017}. More recently, theta EEG NFB, with EEG recorded over the sensorimotor brain areas, was effective to reduce cognitive deficits in elderly patients with MCI, as well as stroke patients \cite{Marlats2020, Mane2020}.

At the state-of-the-art (SENIOR project), "Nudge theory"-based cognitive training can be administered by specialized therapists only, the training level is defined by high-level cognitive assessments (e.g., \gls{mmse} and reaction time tests) at the beginning of the intervention and updated on a regular base. Also, the timing for delivering the nudges, as well as their relevance, is based on the subjective therapist's expertise.
On the other hand, \gls{nfb} training is completely customized and timely adapted to match the user's mental state conditions, via continuous brain monitoring. Also, in the \gls{bci} community, an increasing effort to bring the promising results of \gls{nfb} out-of-the-lab is being ongoing since a long time \cite{Wang2011, Nokia2011, Amaral2017, Alchalcabi2017, Vasiljevic2020}. However, the research on the best strategies for improving learning in \gls{bci} remains a vivid field of investigation.
Thus, SENIOR application of "Nudge theory" could represent a complimentary solution to improve learning in \gls{bci} research. Also, \gls{nfb} could provide contingent intervention that could increase the effectiveness of "Nudge theory"-based cognitive training.

ACTA (Advanced Cognitive Training for Aging) aims to integrate two different domains, the "Nudge theory" and the \gls{nfb} domains, and to get the best from each of them: in fact, ACTA aims to close the loop of a "Nudge theory"-based cognitive training, with introducing an automatic and fully-personalized \gls{nfb} reinforcement based on the instantaneous patient's mental condition.
In particular, ACTA will expand the methodologies of "SENIOR - SystEm of Nudge theory based ICT applications for OldeR citizen" (CARIPLO funded project ID 2018-NAZ-0129, rif. 2018-0826) project, currently ongoing at the University of Milano-Bicocca \cite{SENIOR2019}. Going far beyond SENIOR, ACTA aims to design and test an integrated Nudge-\gls{nfb} cognitive training by introducing a closed-loop assistance based on \gls{bci} and real-time quantitative assessments of the patient's conditions and attention level \cite{Jiray2019,Alegre2003}. 
Moreover, ACTA targets a new low-cost and wireless setup which could be used in less-controlled environments, i.e., in the patient's home, without the need of a clinical caregiver. At the same time, ACTA will provide the clinical caregiver with quantitative outcomes from the patient's behaviour and physiological conditions all throughout the course of the cognitive training.
Thus, ACTA will provide a proof-of-concept that could be, later, easily extended to other use cases (i.e., motor disturbances in Parkinson's disease patients) or cognitive training programs (by adjusting the timings and the complexity of the implemented tasks) \cite{Belkacem2020, Yeo2018, Alchalcabi2017,DYang2019}.

The remainder of this paper is organized as follows: Section\ref{sec:MnM} we present the methodologies to implement ACTA, in Section \ref{sec:results} we disclose the conceptualization of ACTA and the operative choices made in the development of the system and of the experimental protocol. In Section \ref{sec:discussion} we discuss the potentialities, the limitations and the future perspective that ACTA could open for the new generation of tele-medicine and mobile-health. Finally, Section \ref{sec:conclusions} concludes the paper.

\section{Materials and Methods} \label{sec:MnM}

\subsection{Participants}
Forty senior citizens with \gls{mci} from the cohort involved in the SENIOR project will take part in ACTA and will be enrolled via the same clinical and neuropsychological evaluation.
Particularly, body mass index, smartwatch-based physical activity and energy expediture measure, sleep quality, \gls{mmse} score, functional-executive test scores will be obtained for each of them.

The eligibility criteria and the exclusion criteria for ACTA will be the same as in SENIOR, and are recalled here for convenience.
Individuals will be enrolled in ACTA if: (1) they are 65-85 years old, (2) they are diagnosed of \gls{mci} as measured via a standard battery of neuropsychological tests (e.g., \gls{mmse}, (3) they demonstrate an entry level in informatics.
Exclusion criteria are as follows: (1) diagnosis of severe psychiatric disturbances, (2) severe medical conditions that lead to continuous medical assistance, (3) lack of independence in daily activities, (4) motor impairments.

The experimental phases will be conducted in accordance with the Declaration of Helsinki. All experimental protocols and procedures for the first experimental phase are currently under ethics approval by the IRCCS Istituto Auxologico Italiano Ethical Committee. An additional approval will be requested to a Research Ethics Committee for the second experimental phase, which involves the use of neurofeedback via \gls{bci}. However, fully non invasive methodologies are employed (i.e., \gls{eeg}) and the proposed \gls{nfb} approach derives from standard \gls{bci} protocols. In fact, this kind of research has been previously approved in many different countries and many different pathologies \cite{Frontiers2014, Monaco2019}, including \gls{mci} \cite{Yeo2018, Silva2019, Mendoza2018}.
The eligible participants will be fully informed about the purpose, the experimental procedures, the data protection and treatment, the potential benefits and risks of taking part in this experiments. Then, they will be asked to sign a written and informed consent to take actual part in the project.

\subsection{System architecture}
\begin{figure}[h!]
\centering
\includegraphics[width=\textwidth]{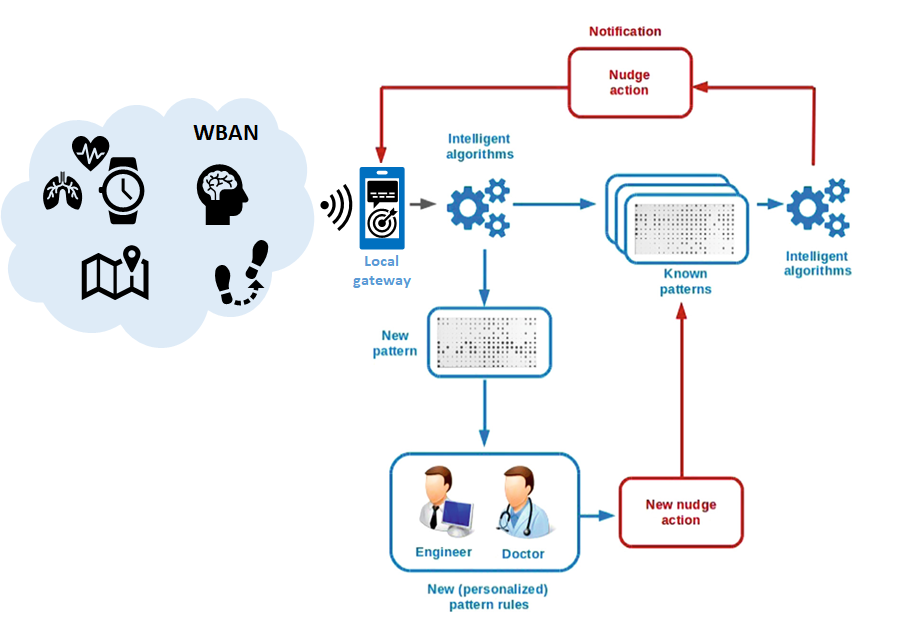}
\caption{ACTA architecture (modified from \cite{SENIOR2019}).}\label{fig:architecture}
\end{figure}

Fig.~\ref{fig:architecture} shows the overall architecture that is being deployed for ACTA. The latter implements the typical three-tier \gls{ioth} architecture \cite{Fragile2020, COMMAG2020, Shukla2019, Fanfan2020, Martiradonna2020, AESM2018} with the first tier including a \gls{wban} for multi-modal sensing \cite{Robel2020,Niknejad2020} to collect physiogical data and vitals from the user (e.g., the heart rate via smartwatch's optical sensor, the brain activity via EEG), as well as the GPS position and the individual's movement (via accelerometer and gyroscope sensors). All data collected are sent through short-range wireless protocols, e.g., Bluetooth, to the smartphone that acts as a local gateway to access the Internet.
At the second tier, the data are aggregated and further process before to reach the cloud. Here, mobile communications, e.g., 3G or 4G, or wi-fi are employed.
The cloud, where data are stored and processed via advanced machine learning to identify individuals' profiles, represents the third tier of the \gls{ioth} architecture.
Finally, the information about profiling are sent back, through the Internet, towards the local gateway to deliver nudges and neurofeedback stimuli, when needed.
From a software point of view, at the front-end, we will extend the SENIOR app to integrate the new functionalities of ACTA (i.e., EEG acquisition, NFB delivery) and the user can run the app on his/her smartphone. At the backend, a \gls{dbms} will store and analyze data on the cloud.

\subsection{Multi-modal wearable sensing design}
ACTA will be used by senior citizens, which are typically less familiar to technology compared to other population segments. Therefore, sensing device, data visualization on smartphone and feedback delivery have to be properly chosen.
In case of senior citizens affected by \gls{mci}, these requirements must be carefully met.

In fact, in ACTA, usability and accessibility, i.e., easy human-machine interaction and minimal intrusiveness \cite{Fragile2020, Mavrogiorgou2020}, are ensured by adopting a fully wearable sensing system, from the heart rate measurement to the brain activity acquisition.

Thus, we adopted the following criteria:
\begin{itemize}
    \item to minimize the sensing intrusiveness;
    \item to minimize the visual modality (except for the delivery of the nudges);
    \item to prefer audio or haptic feedback, such as vibration;
    \item fully wearability, i.e., nnothing to handle with hands.
\end{itemize}

Since we need to get body parameters, i.e., the heart rate, oxygen saturation, and other vitals, a smartwatch is a suitable wearable. Usually, smartwatches come with proprietary software (e.g., Fitbit, Apple) or semi-open software (i.e., Android). To promote inter-operability and reproducibility, we adopted an Android-based smartwatch both in SENIOR as in ACTA.

From the computational point of view, smartwatches are today autonomous computing devices: indeed, they have an operating system, sensors, memory, storage, and connectivity. However, their limited size constrains also the size of RAM and flashdisk, the CPU power and, above all, the battery capacity. For example, a smartwatch has battery capacity in the range of $100mAh$. On the other hand, the GPS consumes about $30mA$. Therefore, the battery may last about  $3h$. Moreover, adding the consumption for the display, the CPU and the connectivity, the device's lifespan could be reduced to less than an hour.

Thus, to increase the system usability and its reliability for longer usages, ACTA system (and SENIOR as well) include both a smartwatch (running Wear OS, a reduction of Android OS) and a smartphone. The smartwatch is solely used for basic sensing of vitals (i.e., heart rate), while the full fledged Android smartphone is employed:  
\begin{itemize}
    \item to act as a local gateway to access the Internet
    \item to synchronize different sensing modalities, including the EEG recordings
    \item to perform basic preliminary computations, before to send the data to the cloud
    \item to host the SENIOR app
    \item to track the individual's position via GPS communications
    \item to deliver nudges and audio feedback
\end{itemize}


Compared to the earlier deployment of SENIOR project, in ACTA the sensing setup is being extended to a low-cost wireless EEG with a limited number of sensors (distributed over the scalp) and the human-machine interaction is being enriched with a real-time neuro-feedback delivery channel.




%

\subsection{Data collection}
The participants will be evaluated at relevant periodic times (at baseline and at different follow-up phases) to assess the effects of the training on their brain activity, motor and functional-executive abilities, as well as to collect information on the usability of the mobile-health system itself.

We will evaluate and collect several behavioral and kinematic measurements: we will exploit GPS-based continuous tracking to decide \emph{when} to provide a nudge or a \gls{nfb} stimulus (i.e., automatically delivered based on GPS position compared to the next landmark). Also, we will measure the path efficiency, i.e., the maximal deviation from the expected ideal path, the peak speed, the reaction time after a new stimulus, the number of steps, and the completion rate, i.e., how many landmarks are successfully reached (similarly to the assessment adopted by \cite{Frontiers2014}). Then, the sensing technologies of the smartphone and the smartwatch allow to collect physiological information, e.g., heart rate, respiration rate, to correlate with behavioural performance.

Also, EEG changes in the brain activity can be investigated through brain network analysis \cite{Sporns2018, Lynn2019}. Particularly, network topology (e.g., modularity), communication efficiency (e.g., small-world), and other graph theory metrics will be extracted from brain networks in real-time (i.e., based on dynamic graph theory) \cite{Dondi2019, Dondi2019b}.

Finally, in line with SENIOR, we will derive a complete evaluation of ACTA usability by administering online questionnaries, available at SENIOR website, about patient engagement, empowerment and quality of experience (e.g., by means of the "Telemedicine Satisfaction Quesitonnnaire" and "Patient Activation Measure and the Patient Health engegnement scale").

\subsection{Tasks}\label{sec:tasks}
The participants will be asked to perform two different training tasks, which combine the use of both motor and cognitive abilities in outdoor path navigation.

\begin{figure}[h!]
\centering
\includegraphics[width=\textwidth]{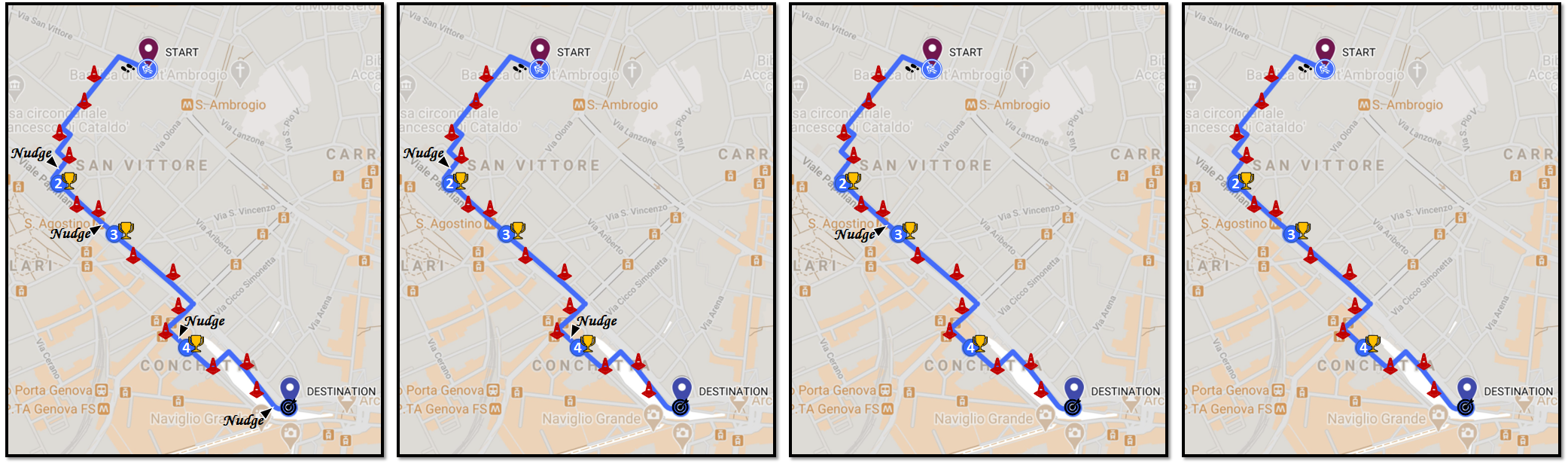}
\caption{SENIOR implementation of "Nudge theory"-based cognitive training.}\label{fig:phase1}
\end{figure}

In task 1, i.e., \emph{memory task}, the patients will be required to achieve a predetermined place (e.g., street, square), within 3 km distance, and asked to repeatedly walk it (in one-way) to memorize it. The path is defined as a pre-determined track from a \emph{start point} to a \emph{final destination}, including a number of intermediate reference locations, namely the \emph{landmarks} (e.g., identified by an incremental index from 1 to 4 in Fig. \ref{fig:phase1}), and a number of other \emph{non-relevant places} (e.g., represented by red cones in Fig. \ref{fig:phase1}). Along the path, appropriate nudges are automatically delivered.

In a first session, consisting of a number of repetitions of the same path, a nudge will be delivered for each landmark (as in the left-most panel of Fig. \ref{fig:phase1}). In the following sessions, the nudges will vanish (in line with the \emph{vanishing cue} strategy), letting the patient finding their own way to reach the expected landmarks (intermediate panels of Fig. \ref{fig:phase1}). Particularly, as the patient's performance increases, the occurrence of nudges will be decreased (or additional support will be provided to complete the path). At the very last session, no intermediate landmarks will be announced using nudges (as in the right-most panel of Fig. \ref{fig:phase1}). This will strengthen and consolidate memory \cite{Amin2016, SENIOR2019}. 


In task 2, i.e., \emph{attentional shift task}, while repeating task 1, the patients will experience an additional \emph{disturbance} (e.g., auditory stimulus asking for answering a question on the smartphone) that will likely provoke an attentional shift. This will train their attention and memory to reach the landmark, despite of the disturbance \cite{DYang2019}.

It is worth mentioning that the timing and the complexity of the path (task 1), as well as the entity of the disturbance (task 2), can be monitored and adjusted to avoid excessive cognitive load or gait fatigue.


\section{Results}  \label{sec:results}

\subsection{Preliminary data} 
Preliminary data are availble from an earlier project carried out by some of the authors of this work at the I.R.C.C.S. Istituto Auxologico Italiano in the framework of the "TEChNology for OBesity" (TECNOB) project (funded by the Compagnia di San Paolo Foundation and supported by technological partners TELBIOS \cite{TELBIOS} and METEDA \cite{METEDA}. TECNOB was a comprehensive two-phase stepped down program enhanced by tele-medicine for the medium-term treatment of obese and diabetic people seeking intervention for weight loss \cite{TECNOB2010, TECNOB2011}.
Seventy-two obese patients with type 2 diabetes have been recruited and randomly allocated to the TECNOB program (n=37) or to a control condition (n=39). 
The TECNOB web-platform supported several functions and delivered many utilities, such as questionnaires, an animated food record diary, an agenda and a videoconference virtual room.
All patients received a multi-sensory armband (SenseWear Pro3 Armband provided by BodyMedia), an electronic tool that enabled automatic monitoring of total energy expenditure (calories burned), active energy expenditure, physical activity duration and levels. They were instructed to wear the device on the back of their upper arm and to record data for $36$ hours every two weeks in a free-living context. Also, the patients could ask for a tele-consultation with the clinical psychologist, to improve self-esteem and self-efficacy, to support motivation, to prevent relapse and to provide problem- solving. 
Within-group analysis showed a significant reduction of the initial weight at all time-points. The median percentage of the initial weight loss for the whole sample was $-5,1$ kg (range from $-6,6$ to $-3,7$) at discharge from the hospital.

\subsection{Low-cost EEG integration}
One contribution of this work is the systematic comparison among the most common portable low-cost \gls{eeg} devices. We operated a selection that is in line with the recent survey presented in \cite{Soufineyestani2020}; however, we focused on portable and low-cost systems, only. A summary of their characteristics is reported in Tab.~\ref{tab:eeg}.


\begin{table}[h!]
\caption{Comparison among commercial low-cost portable EEG devices.\label{tab:eeg}}
\begin{tabular}{l|ccccc}
\hline
\textbf{Brand - Product}	& \makecell{\textbf{Wireless}\\\textbf{(Yes/No)}}	& \makecell{\textbf{Sampling}\\ \textbf{frequency [Hz]}} & \makecell{\textbf{Number of}\\ \textbf{channels}}   & \makecell{\textbf{Noise}\\ \textbf{reduction}}   & \textbf{Price [USD]}\\
\hline
ANT Neuro mini-serie \cite{ANTmini}		& No			& $<2048$      & $8$	& \makecell{\small{active shielding}\\\small{technology for}\\\small{reduction of}\\\small{environmental}\\\small{interference}}       & \makecell{\small{quote}\\\small{needed}}\\
Open BCI \cite{OpenBCI}           		& \makecell{\small{Yes}\\\small{(BLE/WiFi)}}	& $250$         & $8/16/21$	    & n.a.      & $1000$\\
mBrainTrain  \cite{SmartingMobi}     		& \makecell{\small{Yes}\\\small{(BT-EDR)}} & $250/500$    & $24$			& \makecell{\small{high SNR}\\\small{claimed}}       & $66750$\\
Unicorn EEG  \cite{Unicorn}         		& \makecell{\small{Yes}\\\small{(BT)}}		& $250$         & $8$			& \makecell{\small{high SNR}\\\small{claimed}}      & $1200$\\
Wearable Sensing \cite{WearableSens} 		& Yes (BT)			& $300$         & $7-24$			& n.a.      & n.a. \\
Emotiv EPOC-X  \cite{EEPOC}          		& \makecell{\small{Yes}\\\small{(BLE)}}	& $128-256$      & $14$			& \makecell{\small{notch}\\\small{filter}}      & $850$\\
Bitbrain Hero \cite{Hero}       		& \makecell{\small{Yes}\\\small{(BT 2.1+EDR)}}	& $250$         & $9$			& \makecell{\small{active}\\\small{shielding}}   & \makecell{\small{quote}\\\small{needed}}    \\
Brain Live Amp \cite{LiveAmp}  	    	& Yes			& $\leq 1000$         & $8-32$			& n.a.      & $18200$\\
Neurosky - MindWave \cite{MindWave}   & Yes			& $512$         & $1$			& n.a.      & $180$\\
\hline
\end{tabular}
\end{table}

In \cite{comparisonEEG}, a list of brands and products is provided, with further details.

We can notice that all of them are characterized by a low-to-medium sampling frequency (less than $1000$ Hz), a limited number of channels ($21$ max), and a low-to-average price. Additionally, most of them are wireless, but do not have special noise reduction technologies, thus reducing the SNR (i.e., the quality) of the acquired signals. Bluetooth is the most common communications technology used to send the \gls{eeg} samples to the smartphone or the computer wirelessly.
Finally, it is worth mentioning for the most of them the \gls{eeg} sensors are placed in compliance with the International 10-20 System or its extended version, i.e., the 10-10 system \cite{standardEEG1020}.









\subsection{Integration between Nudge and neurofeedback} \label{sec:integration}
The main and most original contribution of this work is the conceptualization of the integration between "Nudge theory" and \gls{nfb}.

One of the most popular paradigms to train a \gls{bci} system for cognitive training is the oddball paradigm, where a sequence of stimuli, with different levels of saliency, are presented to the subject which is required to focus his/her attention to the most salient ones \cite{Rusiniak2013}.
A well-established body of literature reported that the brain reacts to the oddball paradigm producing a specific kind of activity: the so-called event-related potentials (ERPs) \cite{Gu2018}, where different components can be identified as associated to salient (i.e., target) events, or stimuli, and another to non-salient (i.e., non-target) events \cite{Gray2004}. 
The target events are those which the subject is typically instructed to pay attention to, while the others can be regarded as distractors and used as a baseline to identify the attention-related brain patterns.

In SENIOR, the \gls{mci} patient is asked to learn a pedestrian path, walking from one landmark and the next, until destination. The nudges are delivered by cognitive therapists to help the patient reaching out each landmark. Meanwhile, the patient passes by other places, on their way between two consecutive landmarks.
Thus, "Nudge theory"-based cognitive training (as implemented in SENIOR) can be interpreted as an extension of the standard oddball paradigm with application to a more ecological environment: visiting a landmark can be interpreted as a target event, while passing by any other non-relevant place as a non-target event.

\begin{figure}[h!]
\centering
\includegraphics[width=0.7\textwidth]{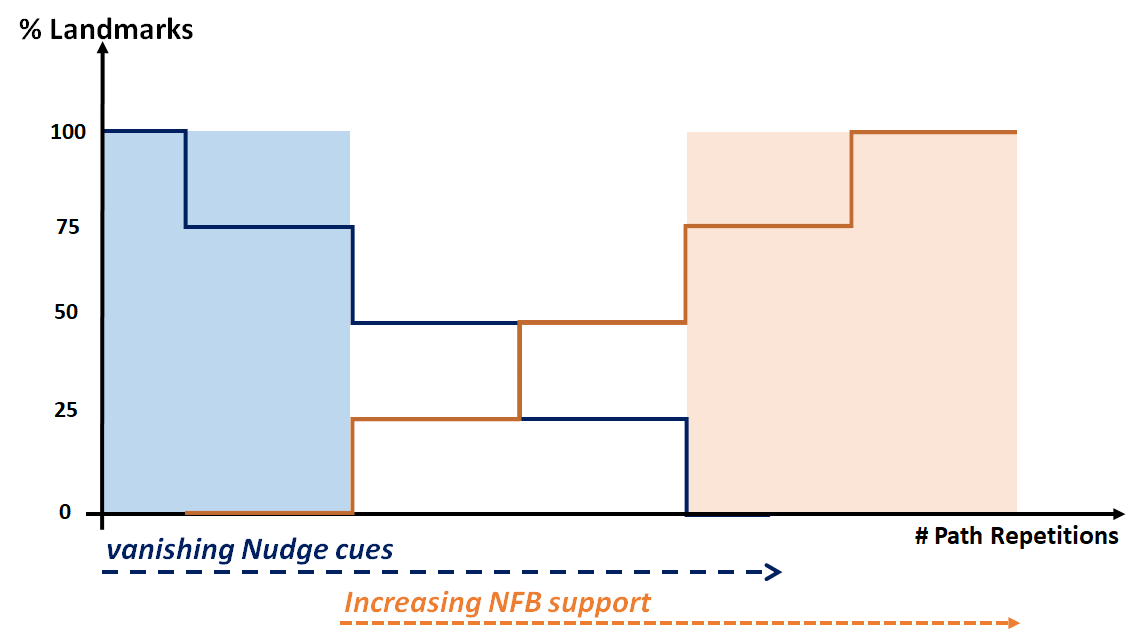}
\caption{An example of outdoor path navigation with presence of nudges and \gls{nfb}.}\label{fig:phases}
\end{figure}

Therefore, based on state-of-the-art (as reported in Section \ref{sec:MnM}), it is fair to hypothesize that closing the loop of SENIOR protocol, i.e., designing a \gls{bci} based on the integration of "Nudge theory" and \gls{nfb}, could improve the beneficial effects of the (open-loop) SENIOR cognitive training protocol.
In the development of such "closed-loop", an "open-loop" training phase has to be planned for allowing the patient learn to control the machine (of the \gls{bci} system), and for the machine to optimize its model of the patient's brain activity.
Therefore, ACTA aims to operate on the two following phases, as shown in Fig. \ref{fig:phases}.

During the first phase, ACTA aims to create a rich training set of \gls{eeg} data (time series or graph metrics) associated with the nudges: in fact, we could extract as many EEG data as many nudges are delivered in the repetitions of the path.
Particularly, if we assume that brain attention can be detected in the proximity of each landmark (as in standard oddball cognitive paradigms), we could label the corresponding EEG data with a positive tag \cite{Jiray2019,YZhang2017}.
Similarly, if we assume that no attentional state is required during the journey between two consecutive landmarks (i.e., passing by the non-relevant places), then a negative label can be associated to the corresponding EEG data. 
Thus, we could exploit advanced supervised ML methods to train a binary classifier that could accurately classify the EEG data either as attention-related or non-attention related samples \cite{Zoppis2020, instaGAT2020, Gadaleta2019, Bressan2020, Kotowski2020}.
At this stage, several deep architectures can be used to identify the brain networks features which are related to the individual's attentional state.
Moreover, by designing and implementating specific "attentional mechanisms", we could weight those pieces of \gls{eeg} information that mostly impact on determining the classification of the individual's attentional level \cite{Zoppis2020, instaGAT2020}.

\begin{figure}[h!]
\centering
\includegraphics[width=\textwidth]{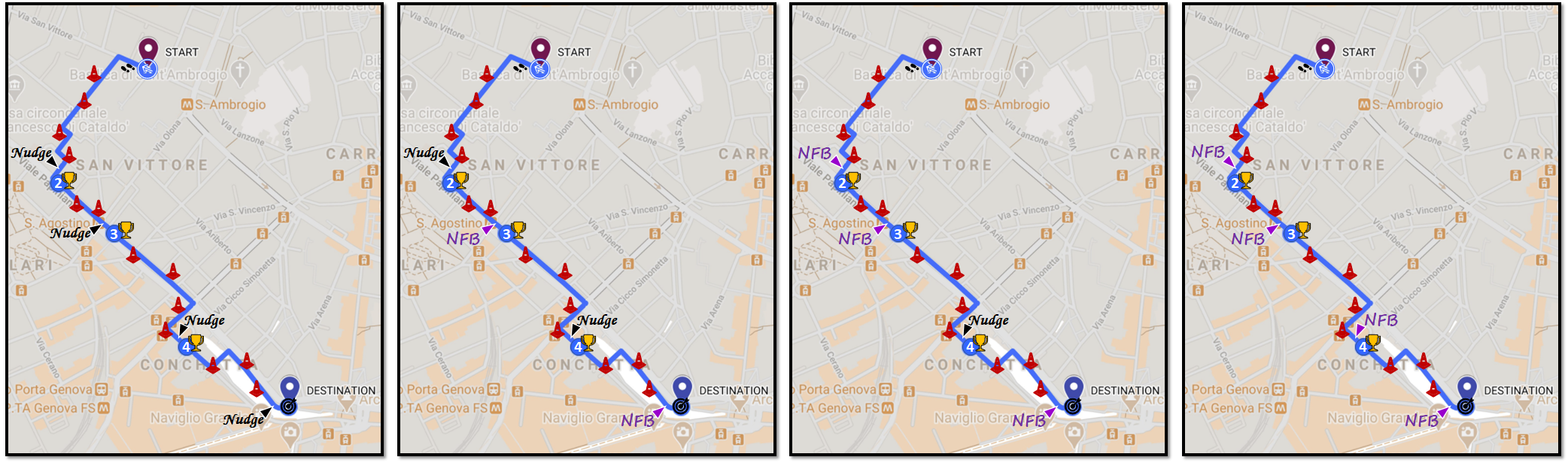}
\caption{Integration between Nudge and neurofeedback.}\label{fig:phase2}
\end{figure}

In the second phase, the same two tasks of the first phase are scheduled (with the patients asked to walk a path of similar complexity as in the previous phase). As sketched in Fig. \ref{fig:phase2}, when the GPS tracking system detects the patients nearby a landmark, either a nudge or a NFB stimulus could be sent him/her through the SENIOR app. In a first session, a nudgeis delivered for each landmark. In the following sessions, the nudges vanish, while the \gls{nfb} is - possibly - delivered every time a nudge is absent. In fact, when a nudge is absent, \gls{nfb} is provided - if and only if - the associated \gls{bci} system could identify an "attention-related" EEG activity (as explained at the beginning of this section).
At the very last session, pure \gls{nfb}-assistance is offered to the patient.
As previously, when achieved the expected location, a "reward" will be sent to the patient.
In task 2, the patients repeats task 1 with the presence of disturbances.

From the analytical point of view, the experimental phase with only nudges (described in Section \ref{sec:tasks}) will represent the training phase to collect samples to train the deep learning models used in the second phase \cite{YZhang2017,Morabito2017}.

Here, the candidate deep model identified during the first phase, and preliminarly trained, subject-wise, over the training set is used to provide the \gls{nfb}.
Also, we assume that the patients are already familiar with the path, thus we expect to classify the EEG activity as "attention", when the patient is nearby a landmark, and as "non-attention", elsewhere \cite{YZhang2017}.
Here, we validate the model by employing it to deliver a \gls{nfb} with the following strategy:
(a) whenever the subject walks nearby a landmark and produces EEG activity that is classified as "attention", we encourage her/him to reach the landmark itself.
(b) whenever the subject walks nearby a non-relevant place and produces EEG activity that is classified as "non-attention", we might provide a positive reinforcement to encourage her/him to carry on her/his own way \cite{Schiatti2018}.
(c) otherwise, i.e., when the classifier's output mismatches the actual patient's location, we might decide to intervene, or not.
When (a) is verified, a positive reward is sent to the patients via the SENIOR app, running on the smartphone.
Labels associated with cases (a) and (b) could be used to update the training set, in line with other recent semi-supervised learning strategies.

\section{Discussion}  \label{sec:discussion}



ACTA represents a new m-health solution that implements an advanced cognitive and motor training based on the integration of the "Nudge theory" and the "neurofeedback theory". Both of them have been previously, and separately, shown to be effective to improve or support the mild cognitive and motor deficits of senior citizens.
Beyond SENIOR, ACTA features a number of new technological strategies that will increase the customization of the therapy, now based on the real-time neural signature of the individual's attention, and the quantitative monitoring of attention and memory abilities all throughout the course of the rehabilitation program. Moreover, ACTA will provide answers to the relationship between such an advanced cognitive and motor training and the brain networks changes \cite{YZhang2017,HYang2020}. This will open up the way to further optimize the training strategies (both in the "Nudge theory" and in the \gls{nfb}).

\subsection{Technological Usability and trustworthiness}
To promote usability and sustainability, we will adapt the software from SENIOR app to acquire data from new sensing modalities (e.g., the \gls{eeg}) and to run \gls{nfb} (i.e., to deliver additional audio or visual stimuli, beyond the nudges).
We keep our sensing devices minimally intrusive, fully wearable, and easy to use \cite{Fragile2020,Healthcom2018,Globecom2018,ICC2020}. Also, we prefer tactile, audio, vibrating feedback (for nudges and \gls{nfb}).
To prolong the battery life in wearables, we decided to use them for basic sensing (i.e., heart rate) and to display stimuli when needed, while the full fledged Android smartphone (running Android OS) is used to access the Internet, synchronize the acquisition from different devices, store data in the cloud, run the SENIOR app, track the individual's position using GPS, provide feedback (as well as distractors of the task 2), as well as phone calls for emergency situations.
Inter-operability is guaranteed by the use of semi-open Android smartwatches, in line with the choice made for SENIOR \cite{Fragile2020, SENIOR2019}. This way we could take advantage of the full range of available Application Programming Interfaces (APIs) in the Android OS (e.g., Firebase cloud sync, Maps API, Text-to-speech API, Phone calls in emergency situations, etc.).
Finally, trustworthiness is ensured as data privacy and protection are designed by fully comply with the GDPR and Italian Legislative Decree no. 196 dated 30/06/2003, across the entire ACTA architecture \cite{Fragile2020, Wowmom2016}.

\subsection{Self-managed therapy and independent aging}
There is an urgent need to understand how technology can be applied in addressing the burden on the healthcare system. In this regards, it should be noted that, the demand for self-management approaches and technologies supporting chronic conditions, e.g., \gls{mci} in senior citizens, is of fundamental relevance.
ACTA aims to contribute to this paradigm shift, by supporting autonomy and independence in elderly with \gls{mci}, via accessible, effective, and easy-to-use ICT platform.
First, we use a low-cost technology to make ACTA accessible to everyone, outside of the advanced research laboratories \cite{Inkjet2019, Niknejad2020,Vasiljevic2020,Alchalcabi2017}. Second, we apply a new rehabilitative approach, based on two different but complimentary training strategies, i.e., the nudges and the \gls{nfb}. They are implemented in ACTA to be fully customized and automatically adapted, in real-time, to the individual's attentional conditions.
ACTA is expected to strengthen the individual's decision-making capabilities (via nudge-based training), and to gradually introduce the self-treatment (based on \gls{nfb}) for the long-term maintenance of his/her cognitive and motor condition.
These choices will allow the elderly to benefit of this kind of cognitive and motor training on a daily basis (i.e., a more intense rehabilitation program can be offered), and to be able to access it in a more ecological environment (e.g., in-home, or outdoor, not necessarily in specialized hospitals only).
On the other hand, ACTA provides the clinicians, who are in charge of the senior citizens using the system, a highly explainable data visualization module to analyze the physiological data of their patients, all throughout the training. This allows the therapist to eventually adjust the program, or to consult with the patients, if needed.

ACTA imposes as a m-health system that can contribute to significantly reduce the cost for the national (i.e., regional) healthcare service, as well as the time spent by patients visiting the hospitals.

\subsection{Toward objective and personalized rehabilitation programs}
Patient-tailored rehabilitation protocols in mental disorders, including \gls{mci}, have become more broadly accepted with the launch of the Research Domain Criteria project in $2008$ \cite{insel2010research}, which was coordinated by the National Institute of Mental Health.
The definition of the protocol and the evaluation of its effectiveness is typically based on the standard classification of mental health given by the \gls{dsmv} \cite{lehman2000diagnostic}, published by the American Psychiatric Association in $2000$. 
However, this categorization has been obtained through questionnaires and by expert observation of the symptoms. In both cases, a high degree of subjectivity could have affected the classification.

Furthermore, a whole spectrum of mental disorders is being demonstrated, i.e., ranging from autism, to schizophrenia and Alzheimer disease. Despite their differences, they could also share similar symptoms. This creates a rather complex and heterogeneous scenario that requires new objective and comprehensive evaluations (i.e., of many aspects of the individual's life and habits) and advanced data analytics competences and tools.
To this regard, ACTA aims to contribute by measuring, in real-time and in a quantitative (yet objective) way the most relevant individual's neuro-physiological features and by adapting the training, accordingly. Indeed, as discussed in Section \ref{sec:integration}, ACTA exploits advanced \gls{ml} to create a mutual learning between the predictive model of the machine and the user who is performing the training. Thus, ACTA is expected to reduce the variability in the \gls{bci} output due to the intra-subject variations, as well as to favour the re-use of the system for other individuals. On the other hand, the two-phase protocol of ACTA (nudges, first, and \gls{nfb} later) is expected to cope with the inter-subject variability and to facilitate the re-use of the system by different users, i.e., producing different brain patterns.

\subsection{Enhancing the understanding of dynamical brain connectivity}
The mathematical and computational tools to characterize the organization and changes of the brain networks (both task-dependent and resting-state) can provide insights into the principles underlying the structure and functioning of the brain (both during tasks and rest). Also, they can help identifying the differences between healthy and pathological individuals \cite{YZhang2017}. However, understanding the causal relationship between a specific neurological disorders and the brain network alterations is still poorly understood.
ACTA aims to contribute to unravel some new insights on the effect mechanisms of human learning by considering advanced \gls{ml} techniques, including some degree of explainability, that allow to explain the association of temporal brain network changes during the training, e.g., learning of a new path \cite{Jiang2017,Schiatti2018}.

These new knowledge could be exploited to further optimize the nudge, or \gls{nfb}, stimulation in order to enhance and speed up the cognitive or motor recovery of patients suffering mental or motor disorders \cite{muldoon2016stimulation}.
This will, in turn, shed some light on how to predict the results of a specific intervention in order to optimize its clinical effectiveness, and to modulate its intensity and target brain regions towards a healthy brain functioning \cite{pasqualetti2014controllability}).

\subsection{Limitations and future perspectives}
To provide fully accessibility to the ACTA mobile-health solution, we included a low-cost commercial EEG device. This kind of systems typically features a low \gls{snr}, can be affected by gait-related artefacts and has a limited number of sensors (less than $20$, as reported in Tab. \ref{tab:eeg}). Unfortunately, signal quality could be critical to ensure an effective training in the framework of ACTA. In this work, we assume that such EEG can provide an SNR high enough to extract attention-related features (as claimed by the vendors). However, in the future, a systematic comparison with research-grade EEG would be useful to validate ACTA \cite{Tavcar2020}.

Also, at present, ACTA exploits cloud computing to run and update the \gls{ml} models that identify the individuals' profiles. In fact, training an advanced \gls{ml} model on a smartphone could be not feasible (especially if the user does not have a very recent smartphone with sufficient computational power). However, real-time processing is needed to run NFB applications, such as ACTA. This is a delay-sensitive healthcare application that could benefit from the presence of an \gls{ai} engine closer to the final user (that uses the service), in order to minimize the delay between ACTA sensing and feedback phases \cite{COMMAG2020, Nguyen2017}. Mobile edge computing and fog computing are seeing increasing popularity as they can lead to minimize network latency and ensure real-time clinical feedback \cite{Nguyen2017, Alnefaie2019, Martiradonna2020}. Finally, network slicing enabled by the new communications generation (5G) could allow to prioritize data transmission of mobile health systems over other delay-insensitive data traffic types \cite{COMMAG2020, Martiradonna2020}.

Finally, profiling is implemented in ACTA using advanced \gls{ml} techniques which guarantee high classification accuracies, but which often lack a high degree of explainability of the classification decisions. However, in mobile-health and in healthcare, explainability and interpretability are highly desirable to provide suitable data visualization and interpretation to clinicians (e.g., to optimize training complexity and rewards), as well as to patients to increase their motivation in the rehabilitation program \cite{Fanfan2020}. In the last few years, a strong effort is being put in the so-called \gls{xai}, which provides strategies and tools to explain the behaviour and the decisions of efficient \gls{ml} models \cite{Lundberg2017,Barus2020, Tjoa2020}.


\section{Conclusions} \label{sec:conclusions}
M-health solutions are seeing an extraordinary development and adoption in a variety of contexts and applications, as a consequence of the ever increasing digitalization of medicine, and a more recent acceleration due to the SARS-CoV-2 pandemics.
This works discloses the conceptualization of ACTA, a new mobile-health solution to support cognitive and motor training in a self-administered way for senior citizens.
Featuring a fully wearable, low-cost, commercial and easy-to-use setup, ACTA represents a promising, fully accessible and highly usable m-health solution for cognitive and motor training or rehabilitation of senior citizens, either in healthy conditions or suffering from mild cognitive or motor impairments.

ACTA has been designed as a minimally intrusive system, promoting a natural human-machine interaction in an ecological environment. Its training program (i.e., the outdoor path navigation tasks) is scalable for the easy administration of new tasks with different complexity, thus representing a valuable tool for clinical psychologists and therapists to support their work. 
It will also provide a user-friendly interface to interact with the senior citizens and make them aware of their progresses during the training program.
Also, ACTA is easy generalized to support different cognitive or motor training for other patients, who might need different training tasks. In fact, ACTA features a data processing module and an advanced \gls{ml} engine that could be adapted to new scenarios. In this case, a new training phase is required for the system to learn the users' profiles and adapt the nudges, and neurofeedback, responses to optimize the training outcomes.

In this way, ACTA pushes forward the recent effort towards a user-centric healthcare, or healthcare 4.0, and imposes as an active aging support for clinicians and senior citizens with mild impairments.

\section*{Acknowledgment}
This work was supported by "SENIOR - SystEm of Nudge theory based ICT applications for OldeR citizen" (ID 2018-NAZ-0129, rif. 2018-0826), a three-year project funded by CARIPLO Fundation in 2018. G.C. was scientific consultant for SENIOR project. A.Z. was granted a scholarship by SENIOR. G.C. was also partially supported by MIUR (Ministry of University and Research) under the initiative Departments of Excellence (Law 232/2016).

\printbibliography

\end{document}